# Spatially Resolved Optical Excitation of Mechanical Modes in Graphene NEMS

*David Miller[1,2,3] and Benjamín Alemán[1,2,3]*

[1]*Department of Physics, University of Oregon, Eugene, Oregon 97403, USA*

[2]*Material Science Institute, University of Oregon, Eugene, Oregon 97403, USA*

[3]*Center for Optical, Molecular, and Quantum Science, University of Oregon, Eugene, Oregon 97403, USA*

**Emerging applications in nanoelectromechanical systems (NEMS) made from two-dimensional (2D) materials demand simultaneous imaging and selective actuation of the mechanical modes. Focused optical probes to measure and actuate motion offer a possible solution, but their lateral spatial resolution must be better than the size of the resonator. While optical interferometry is known to have excellent spatial resolution, the spatial resolution of the focused, laser-based optical driving is not currently known. Here, we combine separately scanned interferometry and optical drive probes to map the motion and forces on a suspended graphene nanomechanical resonator. By analyzing these maps with a force density model, we determine that the optical drive force has a spatial resolution on the order of the size of the focused laser spot. Using the optical force probe, we demonstrate the selective actuation and suppression of a pair of orthogonal, antisymmetric mechanical modes of the graphene resonator. Our results offer a powerful approach to image and actuate any arbitrary high-order mode of a 2D NEMS.**

Nanoelectromechanical systems (NEMS) made from two-dimensional materials, such as graphene[1], h-BN[2], and the transition metal dichalcogenides[3] have high promise for nanomechanical force and mass sensing[4–6] as well as studies of fundamental physics at the nanoscale[7]. Initial experiments with 2D nanomechanical resonators have primarily focused on the dynamics of the fundamental mode[6,8–10], but advanced NEMS applications are increasingly exploiting higher order-mechanical modes[11,12] and the coupling between these modes[13]. For example, by simultaneously tracking several mechanical modes, NEMS resonant detectors can both weigh and localize single molecules or individual viruses[14], while fine control over multiple modes has been used for all-mechanical phonon side-band cooling[13].

Future advances in NEMS multimodal applications demand that the shape of the mechanical modes be precisely known and, simultaneously, that any mode of interest can be efficiently and selectively actuated. Several high-resolution imaging methods, including scanning optical interferometery[15] and atomic force microscopy[16], have already been used to map the mechanical mode shape of 2D NEMS. The fundamental mode and some higher-order modes of 2D NEMS are routinely accessed, but the efficient, selective actuation of a given mode remains a challenge. For instance, a common means to actuate 2D NEMS is with an electrostatic gate[1,5–9,15], but simple gating techniques are inherently inefficient at driving higher-order, antisymmetric modes[13] because the gate applies a symmetric, constant-phase force density across the entire suspended membrane. Furthermore electrostatic gating cannot be used to actuate insulating materials[2] or freestanding 2D drums[17–19] and reduces quality factors[20] due to Joule heating.

Scanning optical interferometry combined with optical drive methods, where an intensity-modulated laser is focused onto the mechanical resonator[21,22], offer an approach to simultaneously image and actuate a 2D NEMS resonator, but only if the optical probe and drive force are sufficiently spatially localized. Optical drive methods can selectively actuate higher-order modes in bulk micromechanical beams because the resulting radiation pressure and photothermal bending forces are localized to the immediate vicinity of the laser spot[23]. When applied to 2D NEMS, however, it is unknown if optical driving can still achieve the same spatial resolution, as there are several reasons that the resolution could exceed the size of the resonator. For instance, 2D NEMS thermalize rapidly because of their ultralow intrinsic heat capacity (mass) and exceptionally high thermal conductivity[24], which can make thermomechanical bending less local. 2D materials also poorly absorb and reflect light, which significantly decreases photon pressure, and they commonly have lateral dimensions ($\sim 2 - 5$ μm) close the size of the laser spot. Nevertheless, optical driving has been employed to actuate 2D NEMS with both defocused (compared to the size of the membrane) and focused lasers[1–3,10,18,24–26]. The defocused drive laser, like gating, exerts a symmetric force and is therefore inefficient at driving higher-order modes. By using a narrowly focused drive laser and scanning it over the resonator, it is possible to infer the spatial resolution of the optical drive. However, experiments to date have either used static lasers, which can only measure mechanical spectra, or co-localized probe and drive lasers[1,10], which convolve spatially resolved motion with actuation and therefore prevent an assessment of the spatial resolution.

To determine the spatial localization of the optical force of a focused laser on a 2D NEMS, we measure the vibrational amplitude of a suspended graphene membrane, which is proportional to the driving force, while we scan the position of a focused, driving laser across the membrane. By comparing the resulting force images to the mechanical mode shapes obtained by scanning optical interferometry, we find that the resolution of optical drive force is limited by the spot size of the laser, and this resolution is sufficient to efficiently and selectively actuate higher-order modes of the graphene membrane.

The graphene NEMS devices we study in this work are few μm diameter nanomechanical drumheads (a 3 μm device is pictured in Fig. 1(a)). We fabricate the drumheads by suspending single-layer graphene over cavities etched into $SiO_2$ on Si using a semi-dry transfer process[27]. The devices are actuated using an amplitude modulated 445 nm laser[1–3,10,24–26], and the amplitude and phase are measured using an interferometer operating at 532 nm and standard lock-in amplifier techniques, similar to previous work[13,15]. A schematic of our experimental setup is shown in (Fig. 1(b)). Both the 445 nm drive and 532 nm probe lasers are scanned using independent dual-axis galvo mirrors and coupled into the same optical path. This allows us to map the mechanical mode shape while driving the drumhead at a specific location. Conversely, we can probe the motion at a specific location, typically an antinode, while scanning the drive across the drumhead. In the following, we present results for a 3 μm diameter device (device 1) and a 5 μm device (device 2), however, we observe similar, reproducible results for other devices and across a range of drumhead sizes.

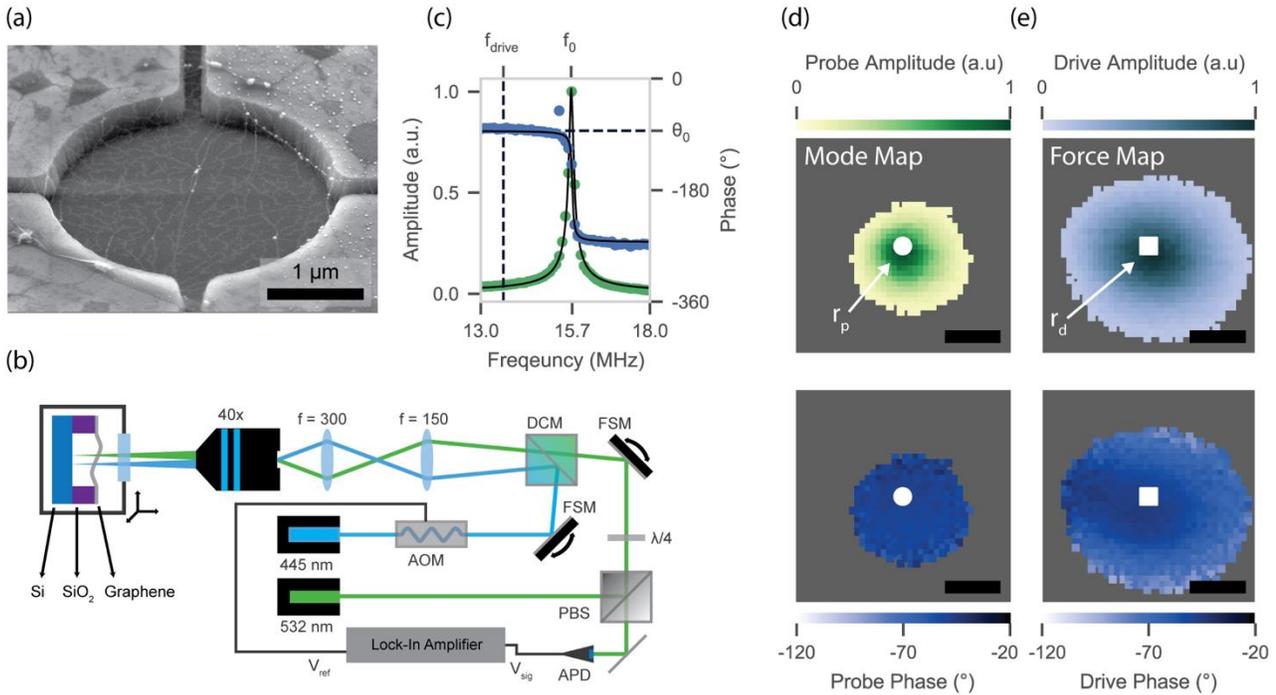

FIG. 1. (a) SEM image of a 3 μm graphene drumhead suspended over a 300 nm cavity. Ventilation trenches allow for air to escape when the device is brought under vacuum. (b) Diagram of optical setup used for measurements. Two scanning mirrors allow for independent scanning of both the drive and probe lasers. λ/4: Quarter Waveplate, AOM: Acousto Optic Modulator, APD: Avalanche Photodiode, FSM: Fast-steering mirror, PBS: Polarizing Beamsplitter, DCM: Dichroic Mirror (500 nm longpass). The 40x, 0.6 NA objective yields a spot-size of ~1 μm for both lasers, which we confirm using the knife-edge trenches in the substrate. (c) Fitted (black lines) amplitude and phase response of the fundamental mode for a graphene drum. The device is driven at a frequency, $f_{drive}$, located below the resonance frequency, during acquisition of the spatial maps. The phase offset, $\theta_0 = -84°$ of the mechanical oscillation, is indicated by the horizontal dashed black line. (d) Amplitude and phase response maps obtained by scanning the probe laser while holding the drive laser at a fixed location, indicated by the white square in 1e. The greyed-out region indicates an amplitude below the noise floor of the detection electronics. (e) Amplitude and phase response obtained by scanning the drive laser while holding the probe laser at a constant position, indicated by the white circle in 1d. (Scale = 1 micron).

We create two types of spatial maps of the membrane: mode maps and force maps. To obtain these maps, first we measure the frequency response spectrum of the graphene drumheads to find the mechanical resonance of the mode of interest. Then, we set the driving frequency well below the mechanical resonance to minimize the variation of the amplitude and phase response due to the position of the drive[15]. Typical amplitude and phase curves with the off-resonant drive frequency indicated are shown in Fig. 1(c). These off-resonant, fixed frequency approach allows scans to be completed in ~1 minute, which is 10-100 times faster than measuring the full frequency response at each point[15]. To obtain the mode map, which measures the membrane's vibrational amplitude at different locations, we fix the position of the drive laser, $r_d = (x_d, y_d)$, (indicated by the white square in Fig. 1(e)) and scan the probe laser over an area slightly larger than the drumhead, while simultaneously measuring both the amplitude and phase at each point of a $40 \times 40$ array, resulting in a 1600 pixel map of the mechanical mode. To obtain the force map, we fix the probe laser position, $r_p = (x_p, y_p)$, (white circle in Fig. 1(d)) on an antinode (*i.e.* a region of maximum amplitude) and we measure the membrane amplitude response, $A(r_d)$, as we scan the position $r_d$ over the membrane. The amplitude of resulting force map is proportional to the drive force through the off-resonant expression $A(r_d) \approx F(r_d)/(m_{eff}\omega_0)$, where $F(r_d)$ is the optical force at position $r_d$, $m_{eff}$ is the membrane effective mass, and $\omega_0$ is the membrane resonance frequency.

The mode and force maps for the fundamental mode of device 1 ($U_{01}$) are shown in Fig. 1(d) and 1(e), respectively. The graphene drumhead used to obtain this data has a resonance frequency of 15.7 MHz and a quality factor of Q ~120. The mode map amplitude (Fig. 1(d), upper) exhibits azimuthal symmetry and it oscillates with a constant phase (Fig. 1(d), bottom) of ~50° across the entire drum (Fig. 1(d), bottom), consistent with theory. The constant phase implies that the relative displacement between the probe and drive lasers does not impact the speed of the mechanical response. The drive force map resembles the measured mode shape (Fig. 1(e), top); it has azimuthal symmetry and its amplitude falls off radially from the center, providing strong evidence that the position of the drive laser does impact the magnitude of the driving force. From the force map, we also see that the membrane responds even when the drive laser is well off the membrane; this has been observed previously[28] and is ascribed to a propagating heat wave. The force phase map is also azimuthally symmetric, but, in contrast to the mode phase map, the phase varies smoothly from $\theta_0 \sim -45°$ near membrane center to $\theta_0 \sim -100°$ at the edge of the drum. The non-constant phase could be due to strain variations or other defects in the drumhead, which cause local changes to the thermal conductivity[26]. Our measurements of the fundamental mode verify the drumhead amplitude response depends on the position of the focused drive laser, and that spatial resolution is less than the membrane size.

To further characterize the optical drive force, we examine the horizontal and vertical polarizations of the antisymmetric degenerate $U_{11}$ mode, which we label $U_{11}^H$ and $U_{11}^V$. The center frequency of the pair is about twice that of the $U_{01}$, with $U_{11}^H$ = 27.4 MHz and $U_{11}^V$ = 31.0 MHz. Both mode maps (see Fig. 2(a)-(b)) show a characteristic antisymmetric shape with two lobes separated by a nodal line[10,15], where one lobe oscillates ~180° out of phase with the other. The amplitude nearly vanishes on the nodal line and the phase change across the nodal line is discontinuous. Positioning the probe laser on an antinode, we again find the drive force maps are qualitatively similar to the mechanical mode maps; they have two lobes separated by a nodal line and a ~180° phase difference across the node. The nodal line of the force map has a location and orientation that is nearly identical to the mode map. The membrane amplitude reaches a near-zero value when the drive laser is positioned on the nodal line, providing strong evidence that the optical drive force is localized to the laser focus.

Although the phase changes by a ~180° across the nodal line in both the mode and force maps, the position-dependence of the phase is quite different for each case. As seen in Fig. 2(c), the mode phase changes abruptly from $\theta$ to $\theta + 180°$, as expected for oscillations that are perfectly out of phase. Surprisingly, however, the force map phase varies continuously across the nodal line by ~0.3°/nm. The continuous phase could result from the force on either side of the nodal line having a relative phase difference that differs from 180°. The total optical drive force is the sum of the force on the left and right antinodes, $F = F_L \cos(\omega t) - F_R \cos(\omega t + \pi + \theta)$. For $\theta = 0$, the phase of the driving force changes by 180° at precisely $F_L - F_R = 0$, which occurs on the nodal line. However, if $\theta \neq 0$, the phase change can be continuous across the nodal line. The non-zero $\theta$ we observe could be due to the thermal propagation time from one antinode to the other, similar to the non-uniform phase seen in Fig. 1(e). We note that the position dependence of the driving force phase could provide a new means to measure the position of a focused laser.

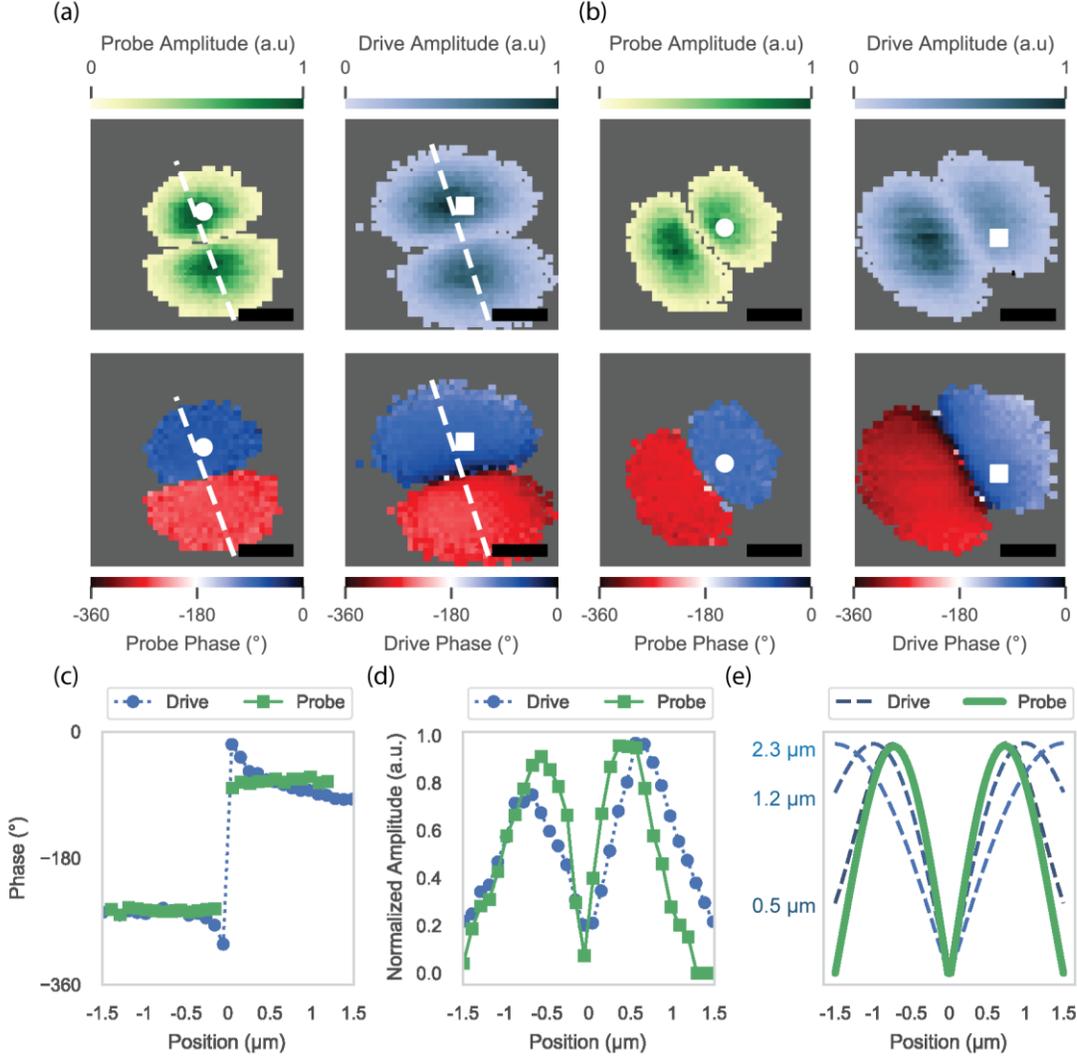

FIG 2: (a) Amplitude and phase recovered while scanning the probe (left) and drive (right) across device 1 with the frequency set just below the resonance frequency for the horizontally polarized $U_{11}^H$ mode. (b) Amplitude and phase while scanning the vertically polarized $U_{11}^V$ mode of device 1. All scale bars are 1 μm. (c). Cross-sectional cut of the phase across the nodal line for both the probe (green) and drive (blue dashed) maps for $U_{11}^H$. The phase change is instantaneous for the probe laser and switches by 0.3°/nm for the drive laser (d.) Cross-sectional profile of the amplitude for the mode and force maps. The transduced amplitude falls to ~25% of its maximum value near the nodal lines. (e) Simulated photothermal force obtained by integrating $U_{11}$ (green solid line) with a gaussian force density at position $r_d$. We plot a cross-section of the calculated force while scanning the drive laser across the nodal for three different gaussian FWHM.

We can infer a measure of the localization of the optical drive force by modeling the optical force density as a gaussian spot and by comparing the force and mode maps. The force exerted on the membrane when the drive laser is positioned at $r_d$ is

$$F(r_d) = \int dA\, U_n(r) f(r; r_d) \tag{1}$$

where $U_n(r)$ is the normalized mechanical mode shape[21–23] and $f(r; r_d)$ is the optical force density. The force $F(r_d)$ is proportional to the experimentally measured force map. We assume that $f(r; r_d)$ is a gaussian distribution of the form $f(r; r_d) \sim \exp\left(\frac{-(x-x_d)^2 - (y-y_d)^2}{2\sigma^2}\right)$, where $\sigma = FWHM/2.355$ gives a characterization of the force localization. A gaussian force distribution will approximate any azimuthally symmetric force centered at the drive laser position, such as photothermal stress or photon pressure. To obtain $F(r_d)$, we numerically integrate the overlap integral in Eq. 1 using $f(r; r_d)$ and the theoretical mode shape of $U_{11}$ for a circular membrane in two-dimensions. Cross-sections of $F(\boldsymbol{r_d})$ perpendicular to the nodal line are shown in Fig 2(e) for various values of σ. For small values of σ (FWHM = 0.5 μm), we find that $F(\boldsymbol{r_d})$ rises rapidly moving away from the nodal line and reaches a maximum before quickly falling off. As σ approaches the width of the membrane (FWHM = 1.2 μm), we find that $F(\boldsymbol{r_d})$ increases slowly from the nodal line and reaches a nonzero value near the membrane boundary. In this case, the gaussian spot is large enough to have significant overlap with the mode shape, even when it is placed near the boundary. We can infer an experimental σ by adjusting σ in the simulated $F(\boldsymbol{r_d})$ to match the measured force and mode maps for the two polarizations of the $U_{11}$ mode. For the $U_{11}^V$ mode profile shown in Fig. 2(d), this process gives an experimental σ between 0.5 and 1 μm, a value approximately equal to the spot size of the focused laser in our setup. Thus, this model predicts that the optical force is localized to the laser position within σ ≈ 1 μm. Though this procedure only yields an approximate value for the localization of the force density, the stark contrast between the measured force map and that predicted from a larger area force density (*i.e.* σ > 1 μm) strongly suggests that the drive force localized to a small region centered around the laser spot. Reducing the laser spot size by using shorter wavelengths or higher NA objectives could further enhance the control of the optomechanical drive efficiency, especially for smaller drumheads or beams which tend to vibrate at high frequencies[29].

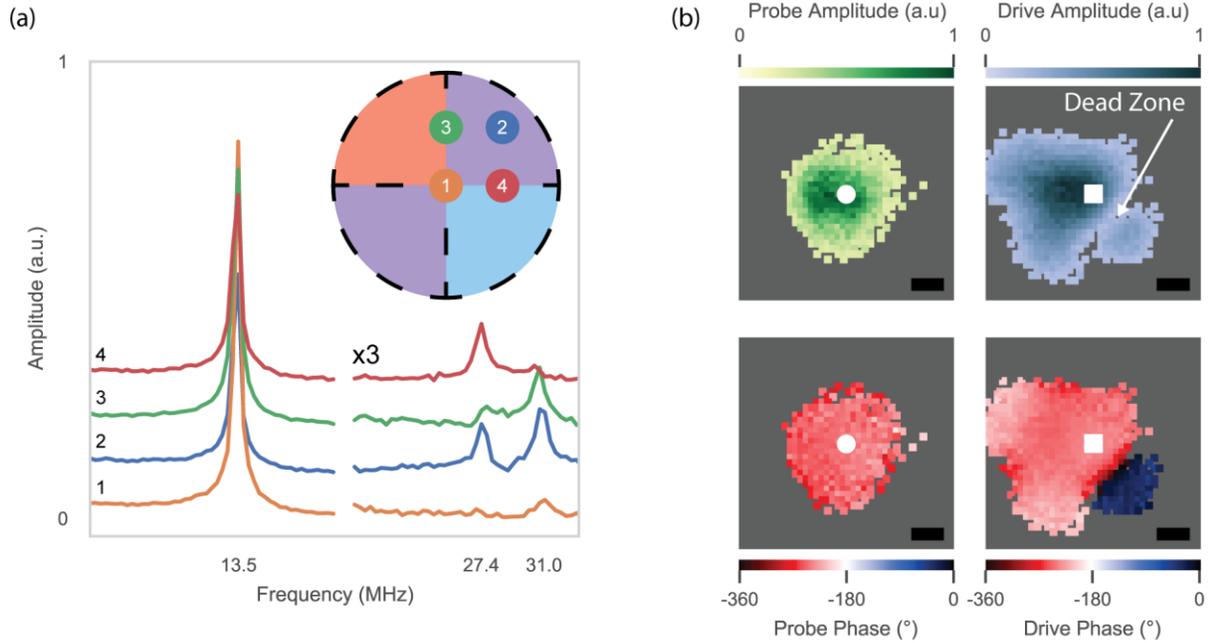

FIG 3: (a) Frequency response traces with the drive laser at four different locations on device 1. 1: A common node of the $U_{11}^H$ and $U_{11}^V$ modes. 2: A common antinode of the $U_{11}^H$ and $U_{11}^V$ modes. 3: A node of the $U_{11}^H$ and antinode $U_{11}^V$ modes. 4: An antinode of the $U_{11}^H$ and a node of the $U_{11}^V$ modes. The probe laser is fixed at location 2 for all measurements as to be sensitive to all three modes. The higher-order modes are scaled by a factor of 3 compared to the fundamental for clarity. (b) Mode and force maps for the fundamental mode of device 2. The force map in this case is more complex than the mode map, unlike those shown for device 1. (Scale = 1 micron).

As further evidence of the local nature of the optical force in 2D NEMS, we demonstrate that a focused drive laser can selectively suppress or excite either polarization of the $U_{11}$ mode. To show this, we position the probe laser at a location sensitive to the motion of $U_{01}$ and both polarizations of $U_{11}$, and then we measure the frequency response while the drive laser is positioned at four different locations on the membrane, all either on an antinode or node of the orthogonal $U_{11}^H$ and $U_{11}^V$ modes (see Fig. 3). The spectra show that the mode can be excited when the drive is placed on an antinode, or suppressed by

~75% when placed on a node. The suppression is sensitive to the beam shape and beam positioning—which we did not fully optimize—so it is possible to achieve a much higher degree of mode suppression. Suppressing individual modes of a degenerate pair is typically quite hard, since they overlap in frequency, making this technique useful for probing the motion of a single mechanical polarization[30]. Placing the drive laser at the point of a mode's maximum response also reduces the need for high laser powers, which can lead to irreversible changes in the device[25]. Although we only study the first three modes here, this technique could also be used at higher frequencies, where the dense spectrum of modes can overlap significantly[31].

In some devices, we observe more complex behavior in the force map than the mode map would indicate. For device 2, the mode map appears to be highly symmetric, similar to theoretical predictions[15] as well as device 1. However, we see significant variation in the phase and amplitude response of the device in the force map and observe a "dead-zone", where the drive force disappears and a 180° phase change occurs (Fig. 3(b)), much like crossing a nodal line. This is a likely due to a breakdown of our assumption of a gaussian force density and could be cause by various defects in the drumhead, such as adlayers or grain boundaries[26]. These observations indicate that the position of the drive force can drastically alter the expected amplitude and phase of a given mode. A better understanding of origin of this non-uniform phase response will be important for experiments which precisely measure the oscillation phase[24].

In summary, we have combined spatially-resolved imaging with a force density model to infer the spatial resolution of the optical drive in a graphene nanomechanical resonator. Despite the fast thermalization, low reflectivity, and micrometer-scale size of the graphene resonator, we found that the optical force is localized to within 1 μm of the laser spot and can selectively and efficiently actuate high-order mechanical modes. The combination of high-spatial-resolution optical drive and read-out enables full multimodal control of suspended 2D nanomechanical resonators for future NEMS applications. Our high-resolution, all-optical approach could be combined with optical beam shaping and spatial light modulation to selectively address an arbitrary subset of resonators within large arrays, a feat not easily achievable with electrostatic gating, or could serve as a point source of propagating mechanical waves for use in 2D nanomechanical circuits[32] and waveguides[33].

## ACKNOWLEDGMENTS

**We acknowledge the facilities and staff from the Center for Advanced Materials in Oregon (CAMCOR), and the use of the University of Oregon's Rapid Materials Prototyping facility, funded by the Murdock Charitable Trust. The authors thank Vincent Bouchiat, Andrew Blaikie, Brittany Carter, Joshua Ziegler, and Rudy Resch for scientific discussions and feedback related to this work. This work was supported by the University of Oregon and the National Science Foundation (NSF) under grant No. DMR-1532225.**